\documentstyle[gsirep]{article}
\newcommand{\E}{{\rm e}}
\newcommand{\I}{{\rm i}}
\newcommand{\D}{{\rm d}}
\newcommand{\oh}{\textstyle\frac{1}{2}\displaystyle}
\title{(Anti-) Deuteron Coalescence in Ultrarelativistic Heavy Ion Collisions}
\author{R. Scheibl and U. Heinz \\
         D-93040 Universit\"at Regensburg }
\date{}
\begin{document}
\maketitle

\noindent
We assume that due to their small binding energy, deuterons from the 
midrapidity region cannot survive 
inside fireballs created in ultrarelativistic nuclear collisions.
All observed deuterons are thus expected to be created from a neutron proton
pair at freeze-out by coalescence. This coalescence process involves small 
relative momenta and is thus a non-relativistic phenomenon. Switching to
center-of-mass and relative coordinates and momenta, the probability
$\D N_{\rm d}(R, P_{\rm d})$
for creating a deuteron with given momentum $P_{\rm d}$ at space-time point 
$R$ is most conveniently evaluated in the center-of-mass frame of this 
deuteron.
The highly relativistic dynamics of the fireball needs to be fully taken 
into account only later on, when calculating the Lorentz invariant deuteron 
spectrum by integrating over the freeze-out hypersurface $\sigma$:
\begin{equation}
E_{\rm d} {\D^3 N_{\rm d} \over \D P_{\rm d}^3} = 
\int P_{\rm d}\cdot \D^3\sigma(R)\ \D N_{\rm d}(R, P_{\rm d}) 
\end{equation}

In non-relativistic statistical quantum mechanics $\D N_{\rm d}$
is calculated by projecting the deuteron 
density matrix on the density matrix of the fireball at freeze-out 
\cite{sato}. The latter contains, however, the relativistic dynamics of the 
heavy ion collision and is difficult to compute. This problem may be 
by-passed by switching to the equivalent Wigner function formalism:

P. Danielewicz has shown in \cite{pd} how, starting from a complete 
quantum mechanical description in terms of density matrices, one can 
via a set of reasonable assumptions reduce the deuteron formation rate
to an integral over a product of nucleon single particle Wigner functions
(which we will approximate by hydrodynamical, local equilibrium distribution 
functions) and a quantum mechanical transition matrix element. The 
transition matrix element implements conservation of energy-momentum in 
deuteron formation via interactions with third particles in the fireball 
and determines the probability of deuteron formation.

We derive the following expression  for 
the number $\D N_{\rm d}$ of deuterons with four-momentum $P_{\rm d}$ 
freezing out at space-time point $R$:
\begin{equation}
\D N_{\rm d}(R, P_{\rm d}) = \frac{3}{(2\pi)^3}\ f_{\rm d}(R,P_{\rm d})\
H^2(R)\ {\cal A}(R,P_{\rm d}) \enspace ,
\end{equation}
where the first term accounts for spin degeneracy and phasespace,
$f_{\rm d}$ is the local thermal equilibrium distribution function
for an elementary particle with deuteron quantum numbers, and
$H$ describes the dependence of the fugacity $\exp(\mu/T)$ 
(and thus of the particle density) on the position in the fireball.
${\cal A}$ implements the quantum mechanics of the formation process and is 
determined by the internal structure of the deuteron and the variation of the
nucleon single particle distribution functions $f$ around the deuteron
formation point $R$:
\begin{eqnarray}
\lefteqn{ {\cal A}(R,P_{\rm d}) =
\int \frac{d^3q\ d^3r}{(2\pi)^3}\  {\cal D}(\vec{r},\vec{q})\ \times
 } \nonumber \\ && \times
\frac{f(+\oh\vec{r},+\vec{q})\ f(-\oh\vec{r},-\vec{q})}
     {f_{\rm d}(\vec{r}=0,\vec{q}=0)}\
\frac{H(+\oh\vec{r}) H(-\oh\vec{r})}{H^2(\vec{r}=0)}
\end{eqnarray}
Here $\vec{r}$ and $\vec{q}$ are the relative coordinates and
momenta of the neutron and proton in the deuteron center-of-mass frame;
${\cal D}$ is the Wigner transform of the internal 
wavefunction $\varphi_{\rm d}$ of the deuteron
\begin{equation}
{\cal D}(\vec{r},\vec{q}) =  \int d^3x\  \E^{ -\I \vec{q}\cdot\vec{x}}\ 
    \varphi_{\rm d}^*(\vec{r}+\vec{x}/2)\, 
    \varphi_{\rm d}(\vec{r}-\vec{x}/2)  \enspace .
\end{equation}

The variation of the local equilibrium distribution functions $f$ is 
due to the hydrodynamic expansion of the fireball. ${\cal A}$
is sensitive to flow gradients in very much the same way 
as the correlation function in two-particle interferometry \cite{csh95}.
In the absence of flow gradients ${\cal A}=1$ for sufficiently large 
fireball volumes; for smaller volumes
one recovers the model proposed by Hagedorn for pp collisions who suggested
${\cal A}= \int_{\Omega} |\varphi_{\rm d}|^2\ dV$ as the probability for 
deuteron production in a small reaction volume $\Omega$ \cite{hagedorn}.

Experimental values for the invariant coalescence factor 
$B_2={\rm d}/{\rm p}^2$ cluster around $(1-2)\cdot10^{-2}$\,GeV$^2$
for heavy ion collisions at the BEVALAC, and are up to a factor 10 smaller
in newer experiments at the AGS and SPS. 

Most recent measurements for $B_2$ come from the CERN Newmass/NA52 
\cite{NA52} experiment for $^{208}$Pb+Pb collisions
at $158\,{\rm GeV}/A$: $B_2 = (1.3\pm0.2)\cdot10^{-3}\,\rm GeV^2$ for 
deuterons and $B_2 = (1.1\pm0.3)\cdot10^{-3}\rm\, GeV^2$ for antideuterons.
To compare numerical results of our model 
with these experimental data, fireball size and flow parameters for
the hydrodynamical description of the reaction zone are needed.  
In our group, efforts are presently being made to determine them 
from single particle and pion correlation spectra. It should be
noted that except for those parameters our model is parameter free;
the deuteron spectrum follows self-consistently from other spectra. 
First estimates for the NA52 data have shown to give the correct order of 
magnitude for $B_2$,
with values for ${\cal A}$ of $\approx 0.6-0.75$ at the space point
with maximal thermal emission.

\end{document}